\documentclass[10pt,aps,prd,nofootinbib,superscriptaddress,twocolumn,preprintnumbers,balancelastpage]{revtex4}
\usepackage{amssymb,amsmath,latexsym,graphics, graphicx,epsfig,multirow,comment,appendix,cancel} 

\usepackage{varwidth,xcolor,afterpage}
\usepackage{color}
\definecolor{Mahogany}{rgb}{0.62,0.24,0.15}
\definecolor{colorLink}{rgb}{0.7,0,0}
\definecolor{colorCite}{rgb}{0,.7,0}
\definecolor{colorURL}{rgb}{0,0,0.7}
\usepackage[colorlinks=true,linktocpage=true,linkcolor=black,citecolor=colorCite
,urlcolor=colorURL]{hyperref}

\newcommand{\V}[1]{{ \bf #1}}
\newcommand{\sun}{\odot}
\newcommand{\earth}{\oplus}

\newcommand{\vmin}{v_\text{min}}
\newcommand{\vesc}{v_\text{esc}}
\newcommand{\Eer}{E_{\text{er}}}
\newcommand{\Enr}{E_{\text{nr}}}

\newcommand{\es}[2] {\begin{equation} \label{#1} \begin{split} #2 \end{split} \end{equation}}

\begin{document}
\title{Modulation Effects in Dark Matter-Electron Scattering Experiments}
\preprint{
MIT-CTP-4712}
\date{\today}

\author{Samuel~K.~Lee}
\affiliation{Princeton Center for Theoretical Science, Princeton University, Princeton, NJ 08544 }
\affiliation{Broad Institute, Cambridge, MA 02142}
\author{Mariangela~Lisanti}
\affiliation{Department of Physics, Princeton University, Princeton, NJ 08544}
\author{Siddharth~Mishra-Sharma}
\affiliation{Department of Physics, Princeton University, Princeton, NJ 08544}
\author{Benjamin~R.~Safdi}
\affiliation{Center for Theoretical Physics, Massachusetts Institute of Technology, Cambridge, MA 02139}

\begin{abstract}
One of the next frontiers in dark-matter direct-detection experiments is to explore the MeV to  GeV mass regime.  Such light dark matter does not carry enough kinetic energy to produce an observable nuclear recoil, but it can scatter off electrons, leading to a measurable signal.  We introduce a semi-analytic approach to characterize the resulting electron-scattering events in atomic and semiconductor targets, improving on previous analytic proposals that underestimate the signal at high recoil energies.  We then use this procedure to study the time-dependent properties of the electron-scattering signal, including the modulation fraction, higher-harmonic modes and modulation phase. The time dependence can be distinct in a non-trivial way from the nuclear scattering case.
Additionally, we show that dark-matter interactions inside the Earth can significantly distort the lab-frame phase-space distribution of sub-GeV dark matter.
\end{abstract}

\maketitle

\section{Introduction}

Weakly Interacting Massive Particles (WIMPs) are one of the most well-motivated candidates for dark matter (DM) and have guided dedicated experimental searches in recent years.  In particular, direct-detection experiments have been optimized to detect neutral, weak-scale $\mathcal{O}(100)$~GeV particles that scatter off nuclei in targets.  Experiments such as LUX~\cite{Akerib:2013tjd}, XENON100~\cite{Aprile:2012nq}, and SuperCDMS~\cite{Agnese:2013jaa} are now so sensitive that they are beginning to probe the highly motivated Higgs-exchange regime.  In the next few years, the sensitivities will improve by several more orders of magnitude, closing in on the most-relevant regions of parameter space for WIMPs~\cite{Cheung:2012qy}.

However, as limits continue to tighten with no hints of signal detection, it is becoming increasingly worthwhile to consider loosening the assumptions on the WIMP paradigm; one possible direction is to consider weakly interacting sub-GeV DM.  Such light DM is motivated by several classes of models, including asymmetric~\cite{Kaplan:1991ah,Kaplan:2009ag,Lin:2011gj}, WIMPless~\cite{Feng:2008ya} and other scenarios~\cite{Boehm:2003ha,Boehm:2003hm,Borodatchenkova:2005ct,Fayet:2007ua,Hooper:2008im,Pospelov:2007mp}.  

For DM candidates below the GeV mass range, elastic nuclear recoil energies fall below current detection thresholds. In this range, the possibility of inelastic processes such as DM-electron scattering leading to ionization become important, because the total energy available to scattering 
is still appreciable.  Direct detection of electron-scattering events was studied in~\cite{Bernabei:2007gr,Dedes:2009bk, Kopp:2009et}.  Recently, there has been a renewed focus on the application of electron scattering experiments to the search for sub-GeV DM~\cite{Essig:2011nj,Graham:2012su}, and the first limits have been set using XENON10 data~\cite{Essig:2012yx}.

Two aspects of DM-electron scattering affect the phenomenology of such signals.  First, the scattering event is inelastic.  Inelastic scattering events have also been explored in the context of nuclear excitations~\cite{Ellis:1988nb}, as well as DM scattering to an excited state~\cite{TuckerSmith:2001hy}.  For an electron scattering event to occur, the minimum velocity that the DM must have to excite the electron depends on the bound-state energy of the electron, which of course depends on the detector target.  As a result, the experiments are not as sensitive to lower-velocity DM, which does not have the requisite minimum energy needed to excite the electron.  

Secondly, the detection rate depends on an ionization form factor, which describes the likelihood that a given momentum transfer results in a particular electron recoil energy.  This form factor can be challenging to calculate, as it depends on the wave function of the scattered electron.  This ionization form factor is target-dependent and shapes the energy dependence of the event rate.  

A significant challenge that will be faced by these experiments is the presence of background.  Identifying unique features of the DM signal that distinguish it from potential sources of background is therefore of the utmost importance.   An annually modulating signal, which arises due to the motion of the Earth around the Sun~\cite{Drukier:1986tm, Lee:2013wza}, is the most prominent example and may be a potentially crucial test of the DM origin of a potential signal.  Characteristics of the modulating signal, such as its phase and fractional amplitude, are expected to depend sensitively on physical parameters such as the assumed DM velocity distribution and the form of the coupling between the DM and target particles. Interactions between the target particle and its surroundings (for example, a target electron in a binding potential) can also have a significant effect. 

As a result of the inelasticity of the signal, combined with the ionization form factor, it is non-trivial to extend our intuition of annual modulation from the nuclear-scattering case to the electron-scattering case.  To address this, this paper presents the first detailed study of the time-dependent rate in electron-scattering scenarios.  We complete a detailed study of the expected modulation fraction, higher-harmonic modes, and expected phase, for both atomic and semiconductor targets.  We find that electron-scattering events typically have large modulation fractions, and that the modulation phase may be affected by gravitational focusing  and local substructure.  

We begin in Sec.~\ref{sec:rate} by introducing a semi-analytic approach to calculating the electron-scattering event rate for both atomic and semiconductor targets, building on previous methods.  Our approach for semiconductor targets should be more accurate than previous analytic approximations to the event rate, such as those in~\cite{Graham:2012su}, while at the same time being more tractable than the full numerical calculations presented in, for example,~\cite{Essig:2011nj}. Sec.~\ref{sec:mod} then applies these techniques to study the time-dependent characteristics of the signal. Sec.~\ref{sec: Earth effects} explores the effects of DM interactions inside the Earth.  We point out that in certain models, DM-nucleus scattering cross sections can be much larger than the DM-electron cross section.  Even though DM-nucleus scattering is not observable directly in the lab for these scenarios because of the low thresholds necessary, DM-nucleus scattering inside of the Earth can modify the lab-frame DM phase-space distribution for large enough cross sections.  We conclude in Sec.~\ref{sec:conc}.

\section{Calculating the Event Rate}
\label{sec:rate}

The kinematics of the inelastic process whereby DM ionizes an atomic electron is more complicated than that of DM-nuclear elastic scattering because the bound electron does not carry a fixed momentum.  As a result, the scattering process may take place with any momentum transfer ${\bf q}$ between the initial and final DM state.  However, when $q = |{\bf q}|$ deviates too far above the inverse Bohr radius, $a_0^{-1} \approx 3.7$ keV, the scattering rate receives a strong wave-function suppression, arising from the fact that it is unlikely for the atomic electron to be found with such a high momentum.  

The relevant momentum transfers are significantly smaller than the nuclear masses we consider, which means that the nuclear recoil energy does not significantly contribute to energy conservation.  As a result, the energy conservation equation reads 
\es{}{
({\bf p_\chi} + {\bf q})^2 = {\bf p_\chi}^2 - 2 m_\chi (\Eer + E_b^i)\,,
}
where $\Eer$ is the electron recoil energy, $E_b^i$ is the negative binding energy of the bound initial state (labeled by the index $i$), $m_\chi$ is the DM mass, and ${\bf p_\chi}$ is the initial DM momentum.  For a fixed $q$, the lowest DM speed $v_\text{min}$ that could induce an electron recoil $\Eer$ is found by taking ${\bf q}$ to be antiparallel to ${\bf p_\chi}$:
\es{}{
v_\text{min} = {q \over 2 m_\chi} + {\Eer + E_b^i \over q} \,.
} 

The count rate for DM-induced electron ionization events is proportional to the average over the DM velocity distribution of the ionization cross section times the DM speed, $\langle \sigma^i_\text{ion} v \rangle$. 
In Ref.~\cite{Essig:2011nj} (see also~\cite{Kopp:2009et}), it was shown that 
\begin{equation}
\frac{d\langle\sigma_{\text{ion}}^{{i}}v\rangle}{d\ln \Eer}=\frac{\bar{\sigma}_{e}}{8\mu^2_{e\chi}}\int dq\,q|f_{\text{ion}}^{{i}}(k',q)|^{2}|F_\text{DM}(q)|^{2}\eta(\vmin,t) \, ,
\label{eq:cross_section}
\end{equation}
where $\mu_{e\chi}$ is the reduced mass of the DM-electron system, and $\eta$  is the mean inverse speed.  The normalized cross section $\bar \sigma_e$ and the DM form factor $F_\text{DM}(q)$ may be calculated from the relevant matrix element for DM-free-electron scattering.  The function $|f_{\text{ion}}^{{i}}(k',q)|^{2}$ is the wave-function suppression factor to ionize an electron in the bound state labeled by $i$ to a final state with momentum $k'$, through a momentum transfer $q$.  We will discuss $|f_{\text{ion}}^{{i}}(k',q)|^{2}$ more later in this section.  However, for now, note that if the final state is a plane wave, then $k'=\sqrt{2m_{e}\Eer}$, where $m_e$ is the mass of the electron.  

The differential scattering rate involves a sum over the differential cross sections for all possible initial electron states, accounting for any degeneracies in the states:
\begin{equation}
\frac{dR}{d\ln{\Eer}}= N_{T}\frac{\rho_{\chi}}{m_{\chi}} F(k') \sum_{i}\frac{d\langle\sigma_{ion}^{{i}}v\rangle}{d\ln \Eer} \, ,
\label{eqRate}
\end{equation}
where $N_T$ is the number of target nuclei and \mbox{$\rho_\chi \approx 0.4$~GeV/cm$^{3}$} is the local DM density~\cite{Catena:2009mf,Pato:2010yq,Bovy:2012tw}.  As in the case of nuclear beta decay, the wave function of the scattered electron is distorted by the presence of the nearby atom, requiring that the rate be corrected by the Fermi factor, $F(k')$.  
In the non-relativistic limit,
\begin{equation}
F(k')= \frac{2\pi\nu}{1-e^{-2\pi\nu}} \, ,
\end{equation}
where $\nu=Z_{\text{eff}}\,(\alpha m_{e}/k')$ and $\alpha$ is the fine-structure constant. $Z_\text{eff}$ is the effective charge that is felt by the scattered electron. Although this is expected to be somewhat larger than unity due to the imperfect shielding of the escaping electron by the remaining electrons, we set $Z_\text{eff} = 1$ throughout. As was shown in~\cite{Essig:2011nj}, this is expected to be a good approximation for outer-shell electrons, while inner-shell electrons may require somewhat higher values of $Z_\text{eff}$. Our choice of $Z_\text{eff} = 1$ is conservative, since larger values are expected to further enhance the rate.

The differential scattering rate depends on the convolution of the atomic physics factor, $|f_{\text{ion}}^{{i}}(k',q)|^{2}$, the particle physics term $\bar{\sigma}_e\,|F_\text{DM}(q)|^{2}$, and the astrophysical input $\eta(\vmin,t)$.  With this factorization in mind, we begin by summarizing the astrophysical input.  
The mean inverse speed
\begin{equation}
{\eta(\vmin,t)\equiv \int_{\vmin}^{\infty}\frac{ f_\oplus \left( \V{v}, t\right)}{v}\,d^3v,}
\end{equation}
depends on the Earth-frame velocity distribution of the DM, $f_\oplus \left( \V{v}, t\right)$, which acquires a time dependence as the Earth orbits the Sun.  
In the Galactic frame, and asymptotically far away from the Sun's gravitational potential, we take the velocity distribution $f_\infty (\V{v})$ to be that of the Standard Halo Model (SHM):
\es{SHM}{
 f_{\infty} (\V{v}) = \left\{ \begin{array}{ll}
{1 \over N_{\text{esc}} } \left( {1 \over \pi v_0^2 } \right)^{3/2} e^{- \V{v}^2 / v_0^2 } \qquad &|\V{v}| < \vesc \\
0 \, \qquad &\text{else} \,,
\end{array}
\right.
}
where $N_{\text{esc}}$ is a normalization factor, and we take \mbox{$v_{0}\approx220$}~km/s~\cite{Kerr:1986hz} and the escape velocity \mbox{$\vesc \approx 550$}~km/s~\cite{Piffl:2013mla}.
\begin{figure}[tb]
   \centering
   \includegraphics[width=3.5in]{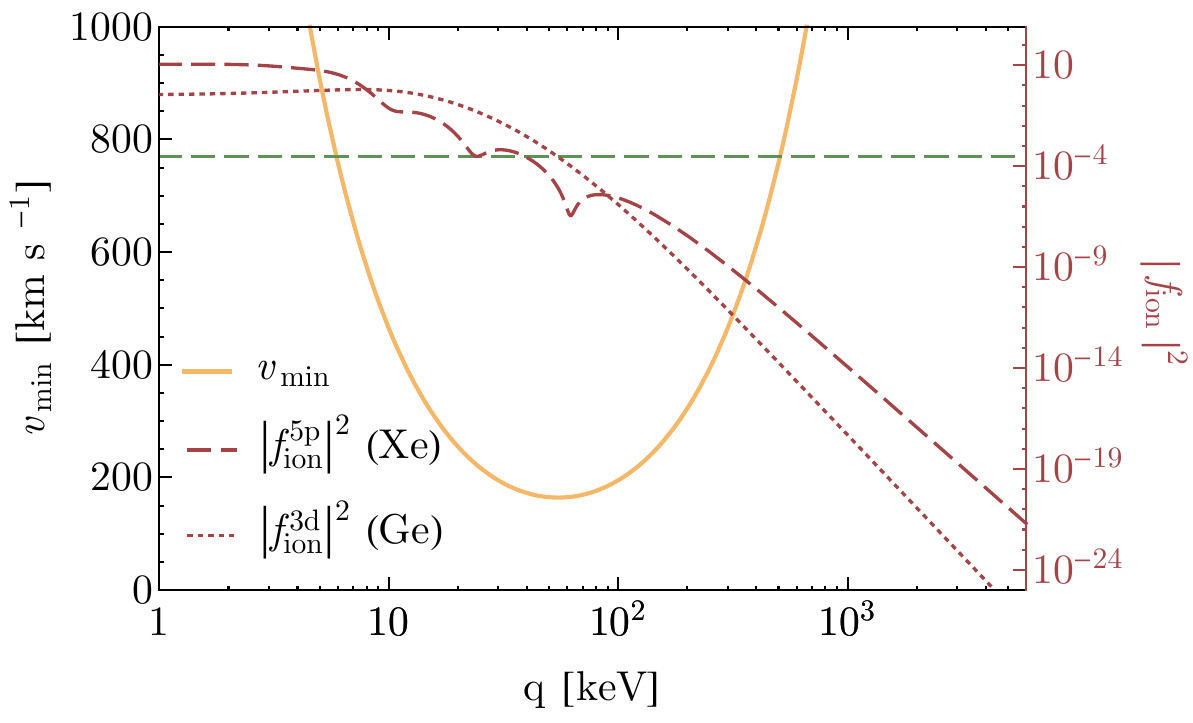} 
   \caption{The integration range in the rate computation includes values of the momentum transfer $q$ for which $\vmin$ dips below the Earth-frame escape velocity ($\vmin \lesssim 770$ km/s, dashed green line).  
At the same time, the ionization form factor $|f_{ion}|^{2}$ (scale on the right) strongly suppresses larger values of $q$. Form factors are illustrated for 100 MeV DM scattering off a $5p$ shell Xenon electron and a $3d$ shell Germanium electron, both with $\Eer = 15$ eV.}
   \vspace{-0.15in}
   \label{fig:qPlot}
\end{figure}

To a first approximation, the velocity distribution at the Earth's location may be found simply by applying a Galilean transformation to $f_\infty (\V{v})$ to transform from the Galactic frame to the lab frame, so that    
\es{eq:Liouville}{
 f_\oplus (\V{v},t) \approx f_{\infty} \left( \V{v}_\sun +  \V{v} +  \V{V}_\earth(t) \right) \,,
}
where \mbox{$\V{v_\sun} = (11, 232, 7)$} km/s~\cite{Schoenrich:2009bx} is the velocity of the Sun in Galactic coordinates and $\V{V}_\earth(t)$ is the time-dependent velocity of the Earth in the Solar frame.  Equation~(\ref{eq:Liouville}) is corrected by the fact that the trajectories of slow-moving DM are deflected in the Sun's gravitational potential~\cite{Alenazi:2006wu,Lee:2013wza}.  This phenomenon, known as gravitational focusing, has important implications for annual modulation, and will be discussed in more detail in Sec.~\ref{sec:mod}.

The particle physics input for the differential rate consists of the normalized cross section $\bar \sigma_e$ and the form factor $F_\text{DM}(q)$.  Following~\cite{Essig:2011nj}, we define 
\es{sigma_bar_e}{
\bar \sigma_e \equiv {\mu_{e\chi}^2 \over 16 \pi m_\chi^2 m_e^2} \left. \overline{ | {\cal M}_{e\chi} (q) |^2} \right|_{q^2 = \alpha^2 m_e^2 } \,
}
in terms of the squared and spin-averaged momentum-space matrix element $\overline{| {\cal M}_{e\chi} (q) |^2}$, so that $\bar \sigma_e$ is equal to the non-relativistic scattering rate between DM and free electrons at momentum transfer $\alpha m_e$.  The form factor $F_\text{DM}(q)$ then captures the $q$-dependence of the matrix element:
\es{FDMq}{
|F_\text{DM}(q)|^2 = \overline{ | {\cal M}_{e\chi} (q) |} /  \overline{ | {\cal M}_{e\chi} (\alpha m_e) |}  \,.
}

For most of this paper, we will frame our conclusions model-independently in terms of $\bar \sigma_e$ and $F_\text{DM}(q)$.  However, it is useful to keep certain models in mind that give rise to specific form factors.  One example is the class of dark-photon mediated models (see, for example,~\cite{Essig:2011nj,Lin:2011gj}) in which the DM is charged under a hidden $U(1)$ gauge group that has a small mixing with the Standard Model photon.  There are two interesting limits in this model: when the dark-photon mass, $m_A$, is much greater than the momentum transfer, and when it is much less.  

To be concrete, let's consider the case where the DM is a Dirac fermion $\chi$.
Then, in the case where $m_A$ is much greater than the momentum transfer $m_\chi v \sim 10^{-3} m_\chi$, the effective Lagrangian for the DM-Standard Model fermion interactions may be written as 
\es{Leff1}{
{\cal L_\text{eff}} = {\epsilon g_D \over m_A^2} \sum_{i}  Q_i (\bar \chi \gamma_\mu \chi)(\bar \psi_i \gamma^\mu \psi_i) \,,
}
where $Q_i$ is the electromagnetic charge of the Standard Model fermion $\psi_i$, $g_D$ is the charge of $\chi$ under the dark photon, and $\epsilon$ is the small mixing parameter between the dark and visible photons.  In this case, $F_\text{DM}(q)  = 1$ and $\bar \sigma_e \approx 16 \pi \alpha \alpha_D \epsilon^2 \mu_{e\chi}^2 / m_A^4$, where $\alpha_D = g_D^2 / 4 \pi$.
The second interesting limit is when $m_A \ll m_\chi v$.  In this case,  $\bar \sigma_e \approx 16 \pi \alpha \alpha_D \epsilon^2 \mu_{e\chi}^2 /  (m_e \alpha)^4$ and $F_\text{DM}(q) = \alpha^{2} m_{e}^{2}/ q^2$.
Motivated by these models, we will consider DM form factors $F_\text{DM}(q)$ = 1 and $\alpha^{2} m_{e}^{2}/ q^2$ throughout the paper; these form factors are generic for heavy and light mediator models, respectively.  

Now, we turn our attention to the ionization form factor $|f_{\text{ion}}^{{i}}(k',q)|^{2}$.  The form factors are calculated differently in atomic and semiconductor targets.  We begin by reviewing the simpler case of atomic targets, and then present an improved semi-analytic approximation for the case of DM-electron scattering in semiconductor targets. 

\subsection{Atomic Target}
\label{Xenon}
\begin{figure}[t]
   \centering
   \includegraphics[width=3.3in]{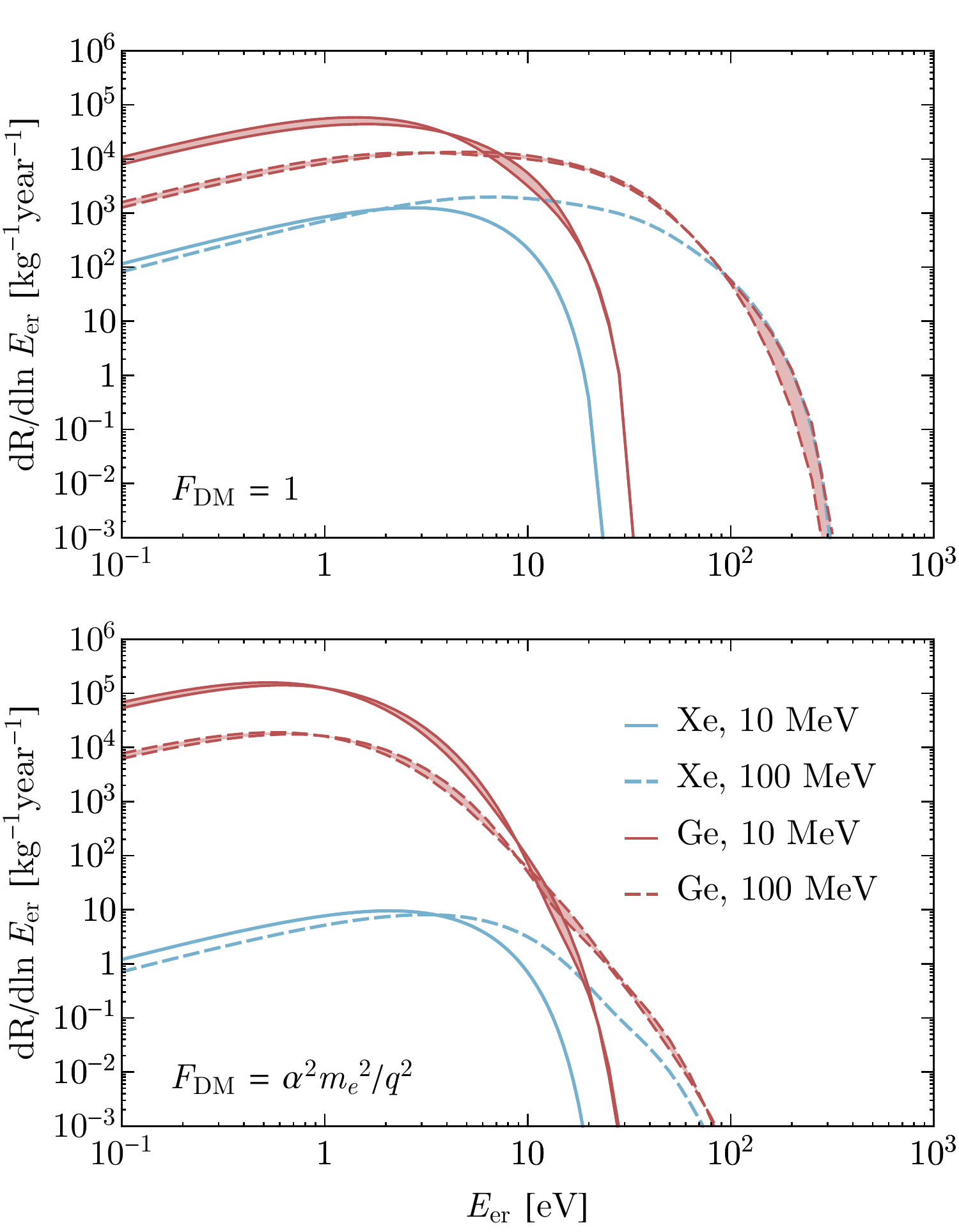} 
   \caption{Differential event rate for DM form factors $F_\text{DM}(q)=1$ (\textbf{top}) and $\alpha^2 m_e^2/q^2$ (\textbf{bottom}), shown for Germanium (red) and Xenon (blue) targets assuming a cross section \mbox{$\bar\sigma_{e}=10^{-37}$~cm$^{2}$} and the SHM.  The solid(dashed) lines correspond to example masses $m_{\chi}=10(100)$~MeV.  The bands in the Germanium curves come from varying the valence-band initial wave functions between $4s$ and $4p$ states. The Germanium lines include the effect of inner-shell $3d$ electrons.}
      \vspace{-0.15in}
   \label{fig:a0}
\end{figure}

The main challenge in calculating the DM-electron scattering rate is determining the ionization form factor, $|f_{\text{ion}}^{i}(k',q)|^{2}$, which depends on the bound-state wave function of the electron, as well as its final-state wave function.  It is most easily determined for the case of an atomic target, so we briefly review this calculation before proceeding to the more challenging case of a semiconductor target.

Assuming plane-wave final states for the scattered electron, so that $k' = \sqrt{2 m_e E_\text{er}}$, the ionization form factor for spherically symmetric full shells with quantum numbers $(n\,\ell)$ reduces to
\begin{equation}
|f_{\text{ion}}^{n\ell}(k',q)|^{2}=\frac{(2\ell+1)k'^{2}}{4\pi^3 q}\int dk\,k|\chi_{n\ell}(k)|^{2} \, ,
\label{atomicFF}
\end{equation}
where $\chi_{n\ell}$ is the radial part of the bound-state wave function and the integral runs over allowed values of $k$ between $|k'\pm q|$ (see, for example,~\cite{Essig:2011nj}).  

As an example, Fig.~\ref{fig:qPlot} shows the ionization form factor for the $5p$ states in Xenon (dashed red) and the $3d$ core-electron states in Germanium (dotted red), 
determined using the Roothaan-Hartree-Fock (RHF) ground-state wave functions and fixing $E_\text{er} = 5$~eV.  The radial RHF wave functions are described as a linear combination of Slater-type orbitals and take the form
\begin{equation}
R_{n\ell}(r) =\sum_{j}C_{j\ell n}r^{n_{j\ell}-1} e^{-\zeta_{j\ell} r} \, ,
\end{equation}
where the coefficients are tabulated in~\cite{Bunge1993113}. Notice that the form factors fall steeply with momentum recoil; 
the ionization form factors strongly bias the scattering towards low-momentum recoil.  In addition, the form factors do not necessarily fall monotonically and thus shape the differential scattering rate.

Also in Fig.~\ref{fig:qPlot}, we show $v_\text{min}$ (solid orange) as a function of $q$ for an example with 100 MeV DM.  When $v_\text{min} \gtrsim 770$~km/s,
no DM is moving fast enough in the Earth's frame to induce ionization, to a good approximation;
for the Xenon example in Fig.~\ref{fig:qPlot},
the allowed momentum transfer is constrained to be $5~\text{keV}\lesssim q \lesssim 500~$keV.
 
Fig.~\ref{fig:a0} shows the differential scattering rate for form factors $F_{\text{DM}}=1$ (top panel) and $F_{\text{DM}}=\alpha^{2}m_{e}^{2}/q^{2}$ (bottom panel) for $m_{\chi}=10$ and 100 MeV assuming a Xenon target.  Only the three outermost orbitals ($5p, 5s$ and $4d$), with respective binding energies $\sim$12, 26 and 76~eV, were used to calculate the rate.  We have verified that the contributions from more tightly bound electrons are negligible.

\subsection{Semiconductor Target}

Next, we consider the case of DM scattering off electrons in a semiconductor target, exciting them above the band gap. Semiconductor materials provide an ideal target to study DM-electron scattering because their band structure allows for electron ionization energies of $\mathcal{O}$(1)~eV compared to noble gas targets with binding energies of $\mathcal{O}$(10)~eV.  For example, any interaction depositing energy above the band gap of $\sim$0.67 eV results in ionization of electron-hole pairs to the conduction band in Germanium.

Several current detector technologies have the potential to take advantage of semiconductor targets to achieve sensitivity to sub-GeV DM.  In these experiments, the scattering signal is  amplified by drifting ionized electrons to induce detectable phonons. The CDMSlite mode of operation of the SuperCDMS experiment~\cite{Agnese:2013jaa}, for example, relies on voltage-assisted amplification of the ionization energy deposited by particle interactions in order to achieve an ionization threshold of 170 eV, which makes it sensitive to sub-GeV DM. Reduction in background levels 
planned for SuperCDMS would further increase the sensitivity~\cite{Sander:2012nia}. The potential to use fully-depleted CCDs to achieve thresholds of 40 eV has been demonstrated by DAMIC~\cite{Barreto:2011zu}.
Given the progress that has been made with the CDMSlite detectors, we will focus mainly on Germanium targets.

Calculating the ionization form factor for a semiconductor target carries with it particular challenges, as the electrons are described by Bloch wave functions in a periodic lattice.  The ionization form factor can be determined using special packages, such as \texttt{Quantum Espresso}~\cite{QE-2009}, as was done in~\cite{Essig:2011nj}.  This is computationally intensive, however, and so an analytic method is also useful for obtaining estimates of experimental sensitivity.  One approach, presented in~\cite{Graham:2012su}, is to model each lattice site in a Germanium crystal by a hydrogen atom with a variable binding energy and calculate the appropriately binding-energy-averaged scattering rate off the $1s$ electron.  
The simplifying assumption of a hydrogenic wave function allows one to obtain analytic expressions for the scattering cross section.  

It is important to understand how well the analytic results in~\cite{Graham:2012su} describe the real ionization form factor in Germanium.  Towards that end, we present an alternate semi-analytic approach here to calculating the ionization form factor, which is related to that in~\cite{Graham:2012su} but relies on using the RHF wave functions for the electrons in Germanium instead of hydrogenic wave functions.  We find significant differences between the results of the two approaches. 

Another difference to keep in mind when considering semiconductor targets lies in the experimental method of detection of a signal. While for atomic targets the final state involves an electron-ion pair, for semiconductor targets it involves creation of electron-hole pairs. 
These final-state charge carriers are drifted using an applied electric field, generating Luke-Neganov phonons.  The energy of the phonons is detected~\cite{Agnese:2013jaa}, giving a direct measure of the number of electron-hole pairs created by the DM scattering.

The number of electron-hole pairs is a function of the total energy $E_d$ deposited into the material by the scattering DM, which is simply related to the electron recoil energy $E_\text{er}$ and the binding energy $E_b$: \mbox{$E_{d}=E_\text{er}+E_{b}$}.   The average energy deposited in order to create an electron-hole pair for Germanium is $\sim$2.9~eV above the band gap.  Thus, we may define the effective number of electrons in the conduction band to be \mbox{$n_{e} = 1 +  (E_d - 0.67 \, \, \text{eV} ) / (2.9 \, \, \text{eV}) $}, taking into account that the initial scattering event promotes one electron from the valence band to the conduction band. We will present results both in terms of $E_d$ and $n_e$.

The electronic states in a semiconductor lattice are described by Bloch wave functions, $\Psi_{\mathbf k}(\mathbf r)$, which may be expressed using Wannier functions:
\begin{equation}
\Psi_{\mathbf k}(\mathbf r) = \sum_{N}e^{{i\mathbf k\cdot\mathbf R_{N}}}\phi(\mathbf r-\mathbf R_{N}) \, ,
\label{bloch_wannier}
\end{equation}
where $\phi(\mathbf r)$ is a Wannier function localized at the site $\mathbf R_{N}$, $\mathbf k$ are the wavevectors in the first Brillouin zone (BZ) consistent with the lattice periodicity, and $N$ is the number of lattice sites.  In the tight-binding approximation, the electrons at a given lattice site are assumed to have limited interactions with the neighboring atoms.  In this case, an atom at a given lattice site is effectively isolated, and the Wannier functions are simply the free atomic orbitals.  Therefore, the Bloch wave function for a given band is the sum over all lattice sites of the associated atomic orbital.   

For our purposes, the expression for the Bloch wave function simplifies even further.  The DM-electron interaction is localized to a single lattice site so long as the momentum transfer is 
\begin{equation}
q \gtrsim (\text{Ge lattice constant})^{-1} \sim~0.4 \text{ keV} \, .
\end{equation}
In this case, the sum over lattice sites in (\ref{bloch_wannier}) disappears and the Bloch wave function is simply the free atomic orbital at the scattering site.  

For large enough momentum transfers, the wave function of the scattered electron can be approximated as a plane wave.  Therefore, the total scattering cross section is obtained by considering the transition of an electron from a localized initial-state atomic wave function--with a ${\bf k}$-dependent binding energy--to a final-state wave function with plane-wave solution, at some energy $E_\text{er}$ above the conduction band minimum.  The atomic scattering can be calculated using the same prescription as that described in Sec.~\ref{Xenon}, with the appropriate RHF wave functions for Germanium.

\begin{figure}[t]
   \centering
   \includegraphics[width=3.3in]{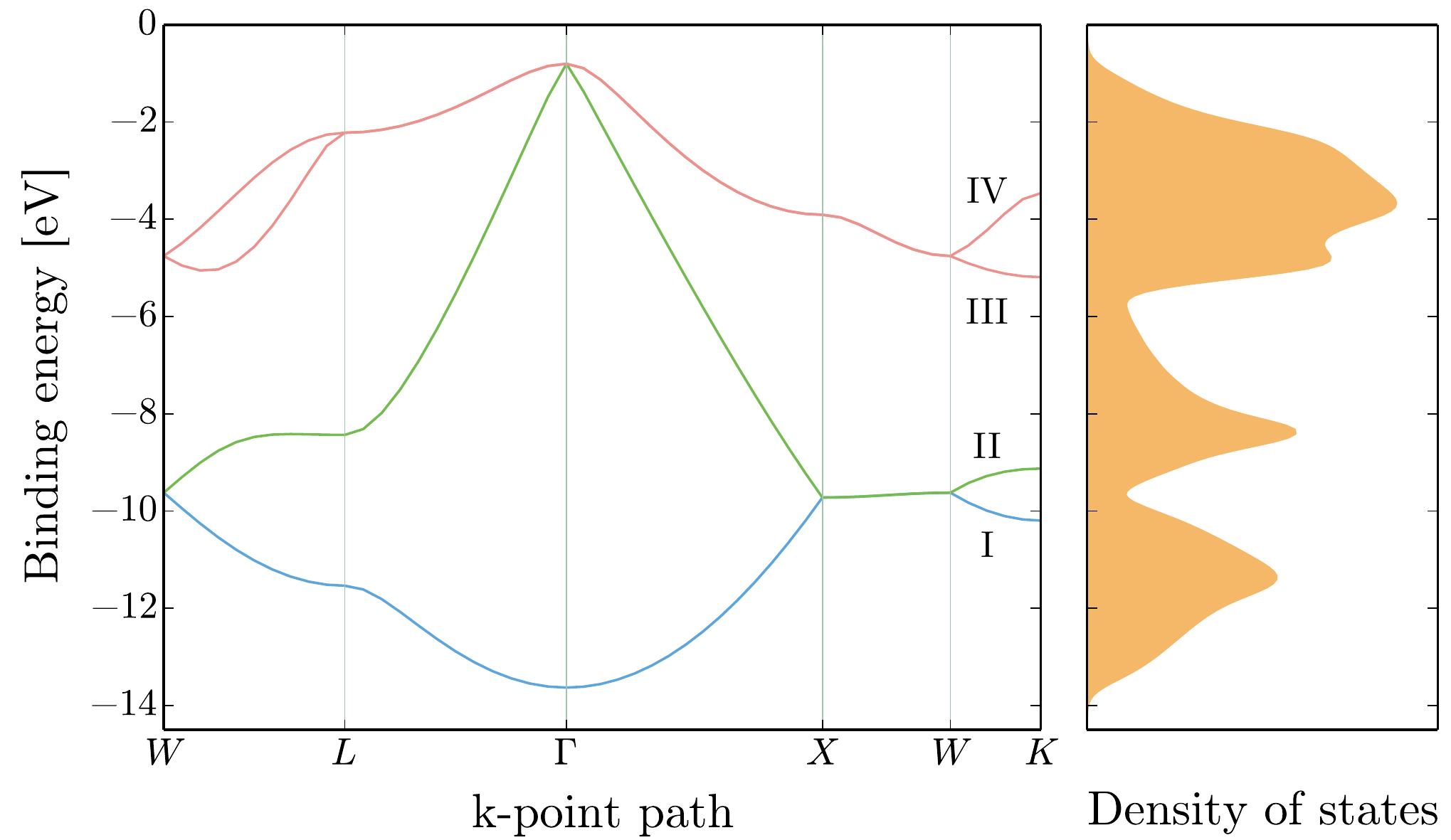} 
   \caption{Band structure of Germanium (\textbf{left}) and the resulting density of states (\textbf{right}) used in the cross-section calculation. Shown is the valence band associated with predominantly $p$-like (III and IV, red), a combination of $s$- and $p$-like (II, green) and predominantly $s$-like (I, blue) states.  The $k$-vectors in the band diagram correspond to a chosen set of high-symmetry points in the first Brillouin zone, with $\Gamma$ being the BZ center. The reference level for the binding energy is taken to be the bottom of the minimum-energy conduction band.}
   \label{fig:dosGe}
      \vspace{-0.15in}
\end{figure}

We are interested in the directionally-averaged rate, for which the variability of the initial bound-state energy $E_{b}$ with ${\bf k}$ may be captured by the valence-band density of states $\rho(E_{b})$ (see, for example,~\cite{Graham:2012su});  the total differential event rate is then obtained by integrating over all binding energies, weighted by the density of states:
\begin{equation}
\frac{dR}{d\ln{\Eer}}\approx N_{T}\frac{\rho_{\chi}}{m_{\chi}} F(k') \int dE_b\,\rho(E_b)\frac{d\langle\sigma_{\text{ion}}v\rangle}{d\ln \Eer} \, .
\label{eq:dR}
\end{equation}
The isotropic valence-band density of states (Fig.~\ref{fig:dosGe}) is computed using the \texttt{GPAW} package~\cite{Mortensen:2005ab}, a density-functional theory code based on the projector-augmented wave method.  For Germanium, the density of states is peaked at bound-state energies of $\sim$4, 8, and 12~eV. These peaks correspond to predominantly $p$-like (III and IV, red), an admixture of $s$- and $p$-like (II, green) and predominantly $s$-like (I, blue) states in the band structure~\cite{Chadi75tight-bindingcalculations}.

It would seem that---depending on the binding energy---we should take a different combination of $s$- and $p$-like atomic wave functions when calculating the expression for $d\langle\sigma_{\text{ion}}v\rangle / d\ln \Eer$ that enters into~\eqref{eq:dR}.  However, we find in practice that taking either the $s$- or $p$-like wave functions independently leads to very similar results for the scattering rate as can be seen from Fig.~\ref{fig:a0}.  As such, we will work with either pure $s$- or $p$-like wave functions for simplicity and estimate our error in this approximation by the difference between the results computed from the respective wave functions.

In addition to the valence electrons, we also consider scattering off the inner-core electrons in Germanium. In general, these electrons do not participate in the bonding process which determines the crystalline structure, with the inter-atomic bonds in Germanium composed of the outer $4s$ and $4p$ states~\cite{Ferry:2013sc}. The least tightly bound set of core electrons reside in the $3d$ shell and can then be accurately described using the appropriate atomic RHF wave function and a shifted binding energy of 30 eV~\cite{Cardona:1978ab}. Including these states can affect the rate, while corrections from subsequent shells are found to be negligible.

We emphasize here that there are many detailed properties of the semiconductor band structure that our calculation ignores.  For example, the energy bands vary along different crystal axes, leading to anisotropy in the density of states.  In addition, depending on the momentum transfer, its effective mass might vary from that of a free electron.  For example, near the minimum of a band, where the curvature is large, the effective mass will be less than $m_e$.  In~\cite{Graham:2012su}, the final-state effective mass was taken to be $m_{*}=0.56\,m_{e}$, corresponding to the density-of-states effective mass at the bottom of the conduction band, for all electron recoil energies $\Eer$. However, if the energy of the electron is sufficiently above the band gap, this effective mass does not accurately describe the final-state dispersion relation.  We assume that the final-state electron is sufficiently above the band gap such that it is well-modeled by a free-electron plane-wave solution, with free-electron mass.  We expect our approximations to break down when the energy of the scattered electron is very close to the conduction band minimum.

Returning to Fig.~\ref{fig:a0}, we can compare the predicted differential scattering rate for a 10 and 100~MeV DM scattering off Germanium with that for Xenon.  For the 10~MeV case, the rate is larger in Germanium for all electron recoil energies; this is a direct effect of the lower binding energy in the semiconductor.  The benefit of the semiconductor target is still present, but less pronounced, for the case of the heavier 100~MeV candidate.  However, the improvement in Germanium's sensitivity over Xenon's is quite dramatic for the form-factor suppressed case in both mass examples.  
\begin{figure}[t]
   \centering
   \includegraphics[width=3.0in]{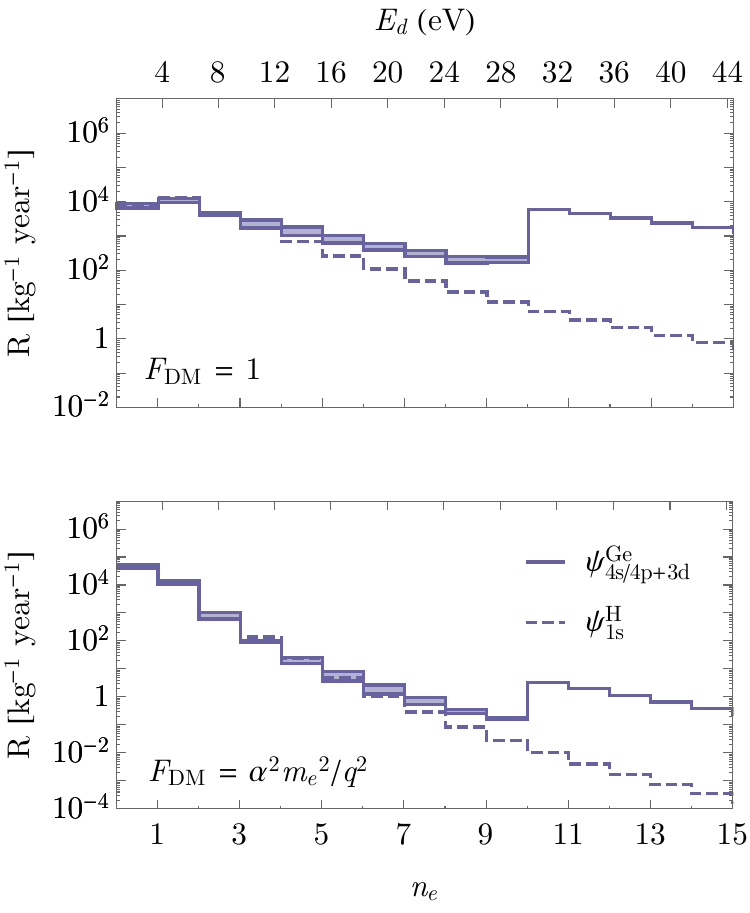} 
   \caption{The binned event rates for $m_{\chi} = 100$ MeV DM obtained using the hydrogenic approach (dashed line) and the Germanium $4s/p$ wave functions (solid band). The latter also includes the effect of inner-shell $3d$ electrons. The differential rates are shown for the cases of a heavy mediator with $F_{\text{DM}}=1$ (\textbf{top}) and a light mediator with $F_{\text{DM}}=\alpha^{2}m_{e}^{2}/q^{2}$ (\textbf{bottom}) assuming $\bar\sigma_{e}=10^{-37}$ cm$^{2}$.  The energy-bin width is that of a single effective electron.}
   \label{fig:comparison}
      \vspace{-0.15in}
\end{figure}

Next, we compare the differential scattering rate calculated using (\ref{eq:dR}) against that obtained by treating each lattice site in a Germanium crystal as a free hydrogenic wave function~\cite{Graham:2012su}.   
By making the assumption that the DM scatters off a $1s$ hydrogenic state ($\psi_{1s}^\text{H}$), it is possible to obtain an analytic solution for the scattering cross section.  From~\cite{Graham:2012su}, the full analytic expression for the $F_\text{DM} = 1$ case is, for example, 
\begin{equation}
\frac{d\sigma}{dE_{\text{er}}}\approx\frac{8\,\bar\sigma_{e}}{3 \pi\alpha\mu_{e\chi}^{2}}\frac{k'F(k')}{v^{2}\,(1+a_{0}^{2}\,q_{\text{min}}^{2})^{3}} \, ,
\end{equation}
with \mbox{$q_{\text{min}} = m_{\chi}v-\sqrt{m_{\chi}^{2}v^{2}-2m_{\chi}(E_{\text{er}} + E_{b})}$} the minimum kinematically allowed momentum exchange. The associated binned scattering rate for a 100~MeV DM is shown in Fig.~\ref{fig:comparison} (dashed blue).
The solid blue band in Fig.~\ref{fig:comparison} shows the range obtained using initial-state $4s$ and $4p$ RHF wave functions for Germanium ($\psi_{4s/4p + 3d}^{\text{Ge}}$).   The $3d$ inner-shell electrons are also included in the Ge calculation and account for the increase in rate at $E_d \gtrsim 30$~eV.  There are clear differences between the two calculations.  Namely, the hydrogenic approach underestimates the number of high-energy scattering events.  Heuristically, this difference can be attributed to outer-shell electrons being on average further away from the nucleus and hence less tightly bound. Using the appropriate $4s/p$ wave functions, as well as the inner-shell electrons, enables us to better model the tail of the differential rate at these energies. 

\begin{figure}[tb]
   \centering
   \includegraphics[width=3.2in]{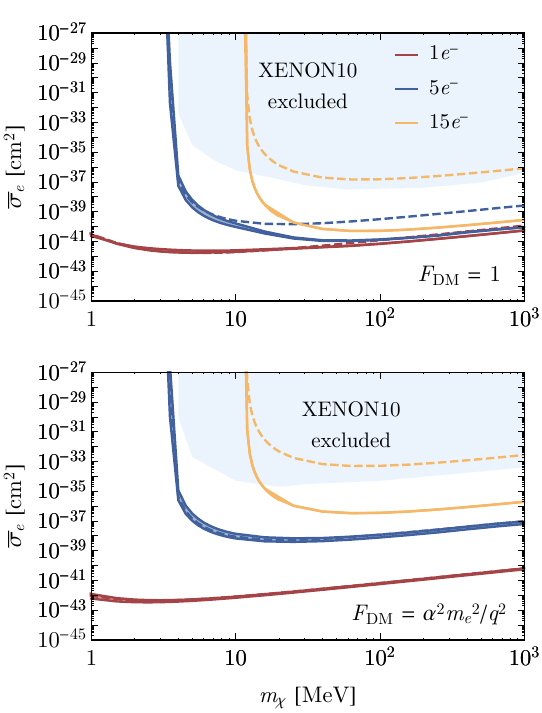} 
   \caption{Sensitivities expected at 95\% confidence level (corresponding to $\sim$3.6 expected events~\cite{Essig:2011nj}) 
   assuming 1 kg$\cdot$year of exposure at a Germanium low-threshold experiment.   These results were obtained using the Germanium RHF $4s/4p$+$3d$ wave functions (solid bands, this work) and the hydrogenic wave functions (dashed lines) for different detector threshold energies, assuming no background.  The thresholds are designated by the number of effective electrons, $n_e$ (1$e^-$, 5$e^-$, 15$e^-$).  The results are shown for the cases of a heavy mediator with $F_{\text{DM}}=1$ (\textbf{top}) and a light mediator with $F_{\text{DM}}=\alpha^{2}m_{e}^{2}/q^{2}$ (\textbf{bottom}).  The XENON10 excluded region at 90\% CL is shown in shaded blue~\cite{Essig:2012yx}.  The hydrogenic approach underestimates the sensitivity at high thresholds.}
   \label{fig:sensitivity}
      \vspace{-0.15in}
\end{figure}

Fig.~\ref{fig:sensitivity} shows the implications of underestimating the high-energy scattering rate on the projected 95\% C.L. sensitivities.  For a low-threshold ($n_e = 1$) experiment, the effect is minimal, as would be expected from the fact that the differential rate for the $\psi_{1s}^\text{H}$ and $\psi_{4s/4p+3d}^{\text{Ge}}$ cases are roughly comparable at low deposited energies.  However, as the threshold energy of the experiment increases, one becomes more sensitive to large $E_\text{d}$, and the differences between the two approximations become much more apparent.  Note that Fig.~\ref{fig:sensitivity} assumes 1 kg$\cdot$year of exposure. 
The shaded region in Fig.~\ref{fig:sensitivity} corresponds to the XENON10 excluded region from the analysis in~\cite{Essig:2012yx}, which was performed with $\sim$15 kg-days of exposure on a Xenon target.

\section{Annual Modulation}
\label{sec:mod}

This section explores the annual modulation of the DM-electron scattering rate.  The time-dependence enters the rate via the Earth-frame DM phase-space distribution $\rho_\chi  f_\oplus(\V{v},t)$, which is determined not only by the velocity of the Earth with respect to the Galactic frame, but also by the position of the Earth in the gravitational potential of the Sun. This latter phenomenon is referred to as gravitational focusing (GF) and is especially important for slower-moving DM particles that linger in the Sun's potential~\cite{Alenazi:2006wu,Lee:2013wza}.

To properly account for GF,~(\ref{eq:Liouville}) must be corrected to include the fact that the trajectories of slow-moving DM are deflected in the Sun's gravitational potential.  In this case, Liouville's theorem requires that
\begin{equation} \label{eq:Liouville1}
\rho_\chi  f_\oplus(\V{v},t) = \rho_\infty f_{\infty} \left( \V{v}_\sun + \V{v}_{\infty} \left[ \V{v} +  \V{V}_\earth(t) \right]\right) \, ,
\end{equation}
where $\rho_\infty$ is the DM density asymptotically far away from the Sun's potential well.  In addition, $\V{v_\infty}[\V{v_{\text{S}}}]$ is the velocity that a particle must have at asymptotic infinity in order to have a Solar-frame velocity $\V{v}_\text{S}$ when it reaches the Earth; it is given by
\begin{equation}  \label{eq:LRL}
\V{v_\infty}[\V{v_{\text{S}}}] = {v_{\infty}^2\V{v}_\text{S} + v_{\infty}(G M_{\sun} / r_{\text{E}}) \V{\hat r}_{\text{E}} - v_{\infty} \V{v}_{\text{S}} (\V{v}_{\text{S}} \cdot \V{ \hat r}_{\text{E}}) \over v_{\infty}^2 + (G  M_{\sun} / r_{\text{E}})-v_{\infty} (\V{v}_{\text{S}} \cdot {\bf \hat r_{\text{E}}}) }\,,
\end{equation}
where $\V{r}_{\text{E}}$ is the position of the Earth in the Solar frame ($\V{\hat r}_{\text{E}}$ is the unit vector and $r_\text{E}$ is the distance between the Earth and the Sun).  Energy conservation requires that \mbox{$v_{\infty}^2 = v^2 - 2  G  M_\sun / r_{\text{E}}$}, with $G$ the gravitational constant and $M_\sun$ the mass of the Sun.  

Gravitational focusing can have a significant effect on the phase of the modulation, as was shown in~\cite{Lee:2013wza} for the case of nuclear scattering, and so we include it in the following analysis, which explores the modulation amplitude and phase of a DM-electron scattering signal in some detail.

\begin{figure*}[tb]
\leavevmode
\begin{center}$
\begin{array}{cc}
   \includegraphics[width=.95\textwidth]{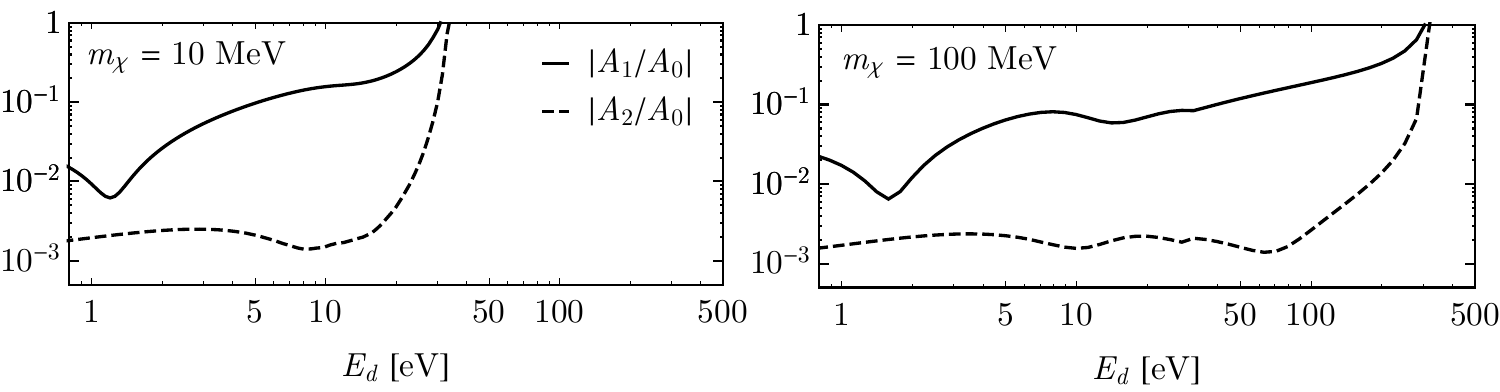} 
 \end{array}$
\end{center}
\vspace{-.50cm}
   \caption{Magnitude of the mode coefficients $A_{1}$ and $A_{2}$ relative to the unmodulated rate $A_{0}$, assuming a Xenon target, for DM masses $m_{\chi}=10$ MeV (\textbf{left}) and 100 MeV (\textbf{right}).  A momentum-independent DM form factor $F_\text{DM}(q) = 1$ is also assumed.}
 	\vspace{-0.15in}  

   \label{fig:harmonics}
\end{figure*}

\begin{figure*}[tb]
\leavevmode
\begin{center}$
\begin{array}{cc}
   \includegraphics[width=.97\textwidth]{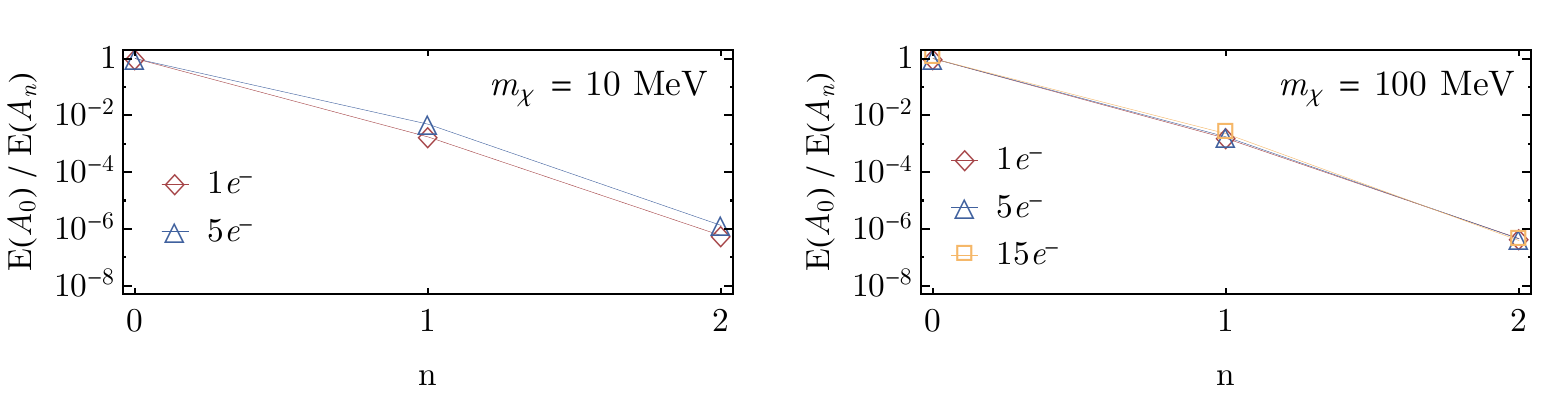} 
 \end{array}$
\end{center}
\vspace{-.50cm}   \caption{Exposure $E$ needed to observe $A_{1}$ (annual modulation) and $A_{2}$ at 95\% significance relative to that needed to observe the unmodulated rate $A_{0}$, assuming a Xenon target and a momentum-independent DM form factor, for DM masses $m_{\chi}=10$~MeV (\textbf{left}) and 100~MeV (\textbf{right}) and various energy thresholds, labeled in terms of $n_e$.}
	\vspace{-0.15in}  

   \label{fig:exposure}
\end{figure*}

\subsection{Modulation Amplitude}

The amplitude of the modulation of the DM-electron scattering rate can be quantified by decomposing the rate into Fourier modes as
\begin{equation}
 \frac{d R}{d E_{\text{er}}}  = A_0 + \sum_{n = 1}^\infty\big[ A_n \cos n\, \omega (t - t_n)\big]\,\,,
\end{equation}
where $\omega$ is the angular frequency of the Earth's orbit.  Furthermore, each $A_n$ is a function of the recoil energy specifying the amplitude of the $n$th mode, which has a maximum at time $t_n$.

The leading mode $A_1$ is conventionally referred to as the annual modulation.  In the case of DM-electron scattering, annual-modulation fractions of ${|A_1/A_0| \sim \mathcal{O}(10\%)}$ are expected and are slightly larger than the $\sim$2--5\% fractions expected in the case of DM-nuclear scattering~\cite{Essig:2011nj,Freese:2012xd}. This results from the strong enhancement of low-$q$ events by the ionization form factor; as demonstrated in Fig.~\ref{fig:qPlot}, scattering events in the energy range of interest are primarily induced by DM from the tail of the velocity distribution (\emph{i.e.}, at large $\vmin$), where the corresponding unmodulated rate $A_0$ becomes relatively small.  
The modes beyond annual modulation can also provide valuable information about the dark sector~\cite{Gelmini:2000dm,Freese:2003na,Savage:2006qr,Fornengo:2003fm,Green:2003yh,Alves:2010pt,Chang:2011eb,Lee:2013xxa}. \
However, because their amplitudes are generally suppressed as $|A_n / A_0| \sim (V_\earth/v_\sun)^n$, detection of these modes typically requires large exposures.

To illustrate these points, we plot in Fig.~\ref{fig:harmonics} the energy dependence of $|A_{1}/A_0|$ and $|A_{2}/A_0|$ for a Xenon target.  We assume the SHM velocity distribution and a momentum-independent DM form factor, and we consider DM masses $m_{\chi}=10$ MeV and 100 MeV.  The amplitude ratios are relatively flat and featureless over most of the relevant energy range, but do increase at high recoil energies that probe the extreme tail of the velocity distribution.  Fig.~\ref{fig:exposure} shows the exposure $E$ (detector mass times measurement period) required to detect the first two modes at 95\% significance relative to that needed to observe the unmodulated rate $A_0$, for various values of the energy threshold, computed as in~\cite{Lee:2013xxa}.  The required exposure grows exponentially with $n$, making detection of the higher modes increasingly difficult.

\subsection{Modulation Phase}

\begin{figure*}[t]
\leavevmode
\begin{center}$
\begin{array}{cc}
   \includegraphics[width=.95\textwidth]{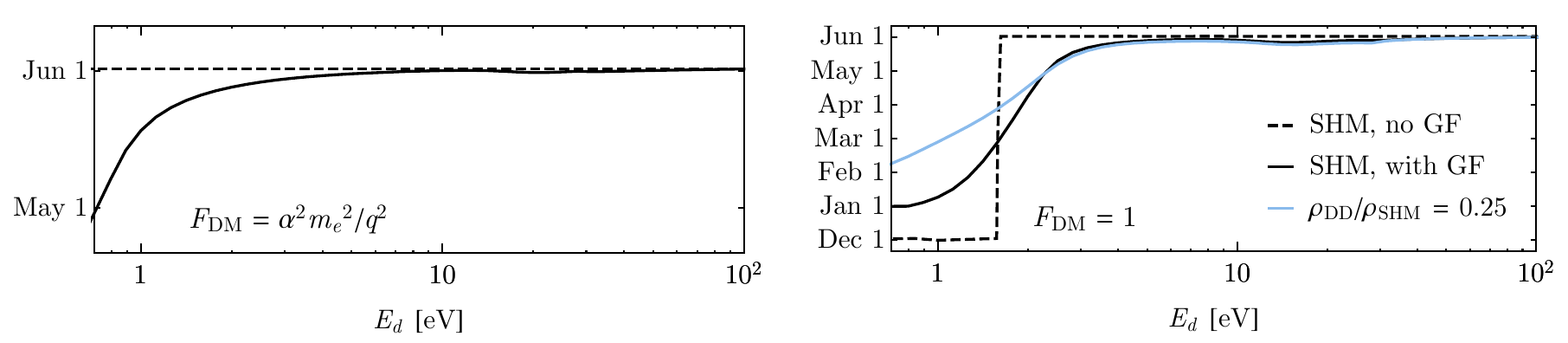} 
 \end{array}$
\end{center}
\vspace{-.50cm}
\caption{Time of year when the event rate is maximized as a function of the total deposited energy in a Germanium target with DM mass of $m_{\chi}=100$ MeV for $F_\text{DM} = \alpha^2 m_e^2/q^2$ (left) and  $F_\text{DM} = 1$ (right).  The solid black curve assumes the SHM velocity distribution, while the blue curve also includes an additional dark-disk component with a relative density of $\rho_\text{DD}/\rho_\text{SHM} \sim 0.25$. For reference, the dashed line indicates the maximum predicted for the SHM when GF is neglected.}
\vspace{-0.15in}
\label{fig:phase}
\end{figure*}

We now turn our attention to the modulation phase, which gives the time at which the event rate is maximized as a function of recoil energy.  Using \eqref{eq:Liouville1} and \eqref{eq:LRL}, the energy-dependent phase can be calculated for a given Galactic-frame velocity distribution $f_{\infty} (\V{v})$.  

We first consider the SHM velocity distribution and review the behavior of the phase of the corresponding DM-nuclear scattering rate.   In this case, the rate is maximized on $\sim$June 1 at high $\vmin \gtrsim 200$ km/s (and hence, at the corresponding high recoil energies).  When GF is neglected, the date of maximal rate abruptly shifts to $\sim$December 1 for $\vmin \lesssim 200$ km/s.  However, when GF is properly accounted for, the date of maximum instead shifts gradually with decreasing $\vmin$, asymptoting to $\sim$December 25 at low $\vmin$~\cite{Lee:2013wza}.  Thus, including GF leads to a shift of nearly a month in the predicted phase at low recoil energies.

The effect of GF is more subtle in the case of DM-electron scattering.  This is because GF more strongly affects slow-moving DM and hence becomes important at ${\vmin\lesssim 200 \text{ km/s}}$; however, as illustrated in Fig.~\ref{fig:qPlot}, events with momentum transfer $q$ corresponding to such $\vmin$ are suppressed by the ionization form factor relative to those corresponding to high $\vmin$.  Despite this, the phase shift (with GF) is similar to the nuclear-scattering case for $F_\text{DM} = 1$, as shown in Fig.~\ref{fig:phase}.  For $F_\text{DM} = \alpha^2 m_e^2/q^2$, the phase shift from June 1 is less enhanced, for the same deposited energy, due to the additional momentum suppression.  

If we relax the assumption of the SHM velocity distribution and consider velocity substructures that increase the proportion of slow-moving DM, the effect of GF on the modulation phase may be enhanced.  For example, simulations find that DM subhalos may be disrupted by the stellar disk of their host galaxy, subsequently merging to form a dark disk that corotates with the stellar disk~\cite{Read:2008fh, Read:2009iv, Purcell:2009yp,Bruch:2008rx}.  These simulations suggest that such a disk might exist in the Milky Way and contribute to the local DM density at the level of $\rho_\text{DD} / \rho_\text{SHM} \sim 0.5$--2, where $\rho_\text{SHM}$ is the non-rotating component of the local DM density.  Other studies suggest that observations may constrain $\rho_\text{DD} / \rho_\text{SHM}$ below this range, however there are significant systematic uncertainties in this result~\cite{Read:2014qva}.  

We shall consider a fiducial case of $\rho_\text{DD} / \rho_\text{SHM} \sim 0.25$.  The dark disk can then be modeled by an additional truncated Maxwellian component, which is boosted in velocity space so that it corotates with the stellar disk and added to the SHM velocity distribution in the correct proportion.  
Taking typical values observed in simulations, we assume a dispersion $v_{0}=70$~km/s and a corotation lag speed $v_\textrm{lag} = 50$~km/s.  The presence of such a corotating component increases the number of slow-moving DM particles in the Solar rest frame, which results in the somewhat larger phase shifts shown in Fig.~\ref{fig:phase}.  It is interesting to compare these results with those predicted for the phase shift in a DM-nucleus scattering experiment in the presence of a dark disk~\cite{Lee:2013wza}.  In that case, it was found that GF can lead to phase shifts at low $v_\text{min}$ of order a month or more.  The same is true here for the DM-electron scattering scenario.  

In this section, we have discussed the modulation of the DM-electron scattering rate resulting from the Earth's motion and GF.  We have examined the energy dependence of the modulation amplitude and phase, highlighting the differences from the case of DM-nuclear scattering that arise due to the different scattering kinematics and the ionization form factor.  We next explore the potential implications of DM interactions inside the Earth on the DM phase-space distribution in the lab frame.

\section{DM interactions inside the Earth}
\label{sec: Earth effects}
   \begin{figure*}[htb]
	\leavevmode
	\begin{center}
	\includegraphics[width=.98\textwidth]{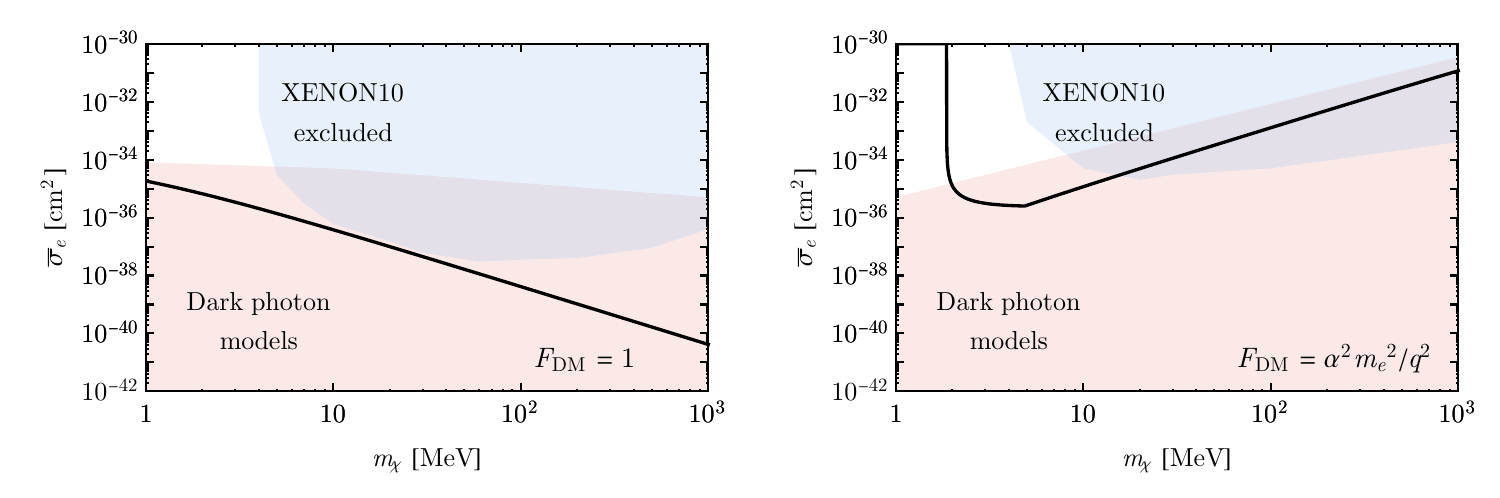}
	
	\end{center}
	\vspace{-.50cm}
	\caption{(\textbf{Left}) The $\bar \sigma_e$-$m_\chi$ parameter space in the dark-photon mediated DM model with DM form factor $F_\text{DM}(q)=1$.  For cross sections above the black curve, we estimate that the DM will scatter at least once off of nuclei with a large angle while traversing the Earth's interior.  (\textbf{Right}) The same as the left panel, but taking DM form factor $F_\text{DM}(q) = (\alpha m_e / q)^2$.  The XENON10 excluded regions~\cite{Essig:2012yx} are shown in blue and the parameter space for the dark-photon mediated models~\cite{Essig:2012yx} is indicted by orange.  }
	\vspace{-0.15in}  
	\label{Fig: Fq}
\end{figure*}

The DM phase-space distribution near the surface of the Earth can be distorted if the DM scatters with nuclei while traversing the Earth's interior.  In the lab frame, these distortions may even acquire time dependence as the Earth rotates, leading to a daily modulation.  A daily modulation induced by scattering in the Earth's interior has been discussed before in the context of nuclear recoils by~\cite{Cline:2012is,Collar:1993ss,Foot:2003iv,Hasenbalg:1997hs,Kouvaris:2014lpa,Mack:2007xj,Sigurdson:2004zp,Zaharijas:2004jv}.   

A key feature that experiments searching for DM-induced ionization signals should keep in mind is that, even though the DM may be so light that the DM nuclear recoils are undetectable in the lab, the nuclear recoil cross section may be significantly larger than the ionization cross section.  This opens the possibility that the light DM may have nuclear recoils inside the Earth before being detected in the lab by electron ionization. If the nuclear recoil cross section is strong enough, this effect can significantly alter the lab-frame DM phase-space distribution.  

The dark-photon model discussed in Sec.~\ref{sec:rate} yields DM-nucleon scattering at tree level, in addition to DM-electron scattering.  We now estimate the cross sections where Earth effects are expected to become important in this scenario, for the mass range $m_e < m_\chi < m_N$, where $m_N$ is the mass of the nucleus. Note, however, that this model merely serves as an illustration, and  the results also apply to other DM models with similar low-energy physics.

In the case where the dark-photon mass $m_A$ is much greater than the momentum transfer $m_\chi v \sim 10^{-3} m_\chi$, the differential cross section for nuclear recoils is
\es{dsigdEr}{
{d \sigma_{N \chi} \over d \Enr} \approx {8 \pi Z^2 \alpha \alpha_D \epsilon^2   m_N \over m_A^4 v^2} \,,
}
where $Z$ is the atomic number of the nucleus and $\Enr$ is the nuclear recoil energy. The atomic number enters instead of the mass number because the DM only interacts with the protons.

The recoil energy is related to the scattering angle $\theta$ in the center-of-mass of the DM-nucleus system by 
\es{}{
\Enr = {2 \mu_{N \chi}^2 \over m_N} v^2 \sin^2(\theta/2) \,,
}
where $\mu_{N \chi}$  is the reduced mass of the DM-nucleus system.  The total cross section is obtained by integrating~\eqref{dsigdEr} over all allowed recoil energies.  
  It follows that the total nuclear-scattering cross section obeys the relation 
\es{echi}{
{\bar \sigma_{e} \over \sigma_{N\chi}} \approx \left({ \mu_{e\chi} \over Z \, \mu_{N \chi}} \right)^2 \,,
}
where ${\bar \sigma_{e}}$ is the electron-scattering cross section for the heavy mediator case.  Importantly, $\sigma_{N\chi}$ is much greater than $\bar \sigma_e$ for typical elements inside the Earth and for DM masses between the electron and nuclear mass scales. 

Now consider a DM particle traveling through the Earth.  There is some probability that this DM particle will have a single scatter with nuclear matter.  Because the differential cross section is isotropic in the center-of-mass frame---and the center-of-mass frame almost coincides with the lab frame---there is a high chance that after the scattering event, the DM will recoil at a large angle.
To estimate the number of large-angle scattering events, we model the Earth simply as a uniform sphere comprised of approximately 32\% Fe, 30\% O, 15\% Si, and 14\% Mg, ignoring all other trace elements. 
Given that the Earth has a density $\sim$$5.5$ g/cm$^3$ and radius $\sim$$6\times10^3$~km, the DM-electron scattering cross section must satisfy
\es{IE1}{
{\bar \sigma_e } \gtrsim & \,\, 4 \times10^{-39} \left({ 100 \, \, \text{MeV} \over m_\chi} \right)^2 \left( {\mu_{e\chi} \over 0.5 \, \, \text{MeV}} \right)^2 \, \, \text{cm}^2\, , \\
}
in the limit $m_\chi \ll m_N$, in order for there to be at least one scattering event as the DM traverses the Earth's interior.  This bound is illustrated in the left panel of Fig.~\ref{Fig: Fq}; Earth effects are important for $\bar \sigma_e$ above the black curve.  The XENON10 excluded region (shaded blue) and the allowed parameter space for the dark-photon mediated model (shaded orange) are also shown~\cite{Essig:2012yx}.  As Fig.~\ref{Fig: Fq} illustrates, there is a clear region of allowed parameter space where Earth effects should be important.\footnote{In specific scenarios, collider constraints from \emph{e.g.} monojet searches may also be constraining~\cite{Khachatryan:2014rra, Aad:2015zva}.}
  
Note that~\eqref{IE1} is specific to a dark-photon model where the DM has a tree-level coupling to quarks.  If the only Standard Model fermion $\psi$ appearing in~\eqref{Leff1} is the electron, then the nucleon-DM coupling arises at one loop~\cite{Kopp:2009et}.  In this case, the right-hand side of~\eqref{IE1} should be divided by a factor $\mathcal{O}(\alpha^{2})$, which significantly reduces the allowed parameter space for Earth effects in Fig.~\ref{Fig: Fq}.

Next, we consider the limit where the momentum transfer $q$ is much greater than the dark-photon mass $m_A$.  In this case, the differential cross section for DM-nucleon scattering is
\es{diff1}{
{d \sigma_{N \chi}  \over d \Enr} \approx {2 \pi Z^2 \alpha \alpha_D \epsilon^2 \over v^2 m_N \Enr^2} \, .
}
Importantly, the total cross section is IR divergent, which results from the fact that there is a new long-range force.  However, since the electrons are also charged under the dark photon, we expect the force to be screened over a distance $\sim$$a_0$, just like in ordinary electromagnetism.  We take this into account by requiring\footnote{In these DM models, the DM-electron ionization scattering events are also screened for momentum transfers \mbox{$q \lesssim a_0^{-1} \approx 4$}~keV.  This may affect the detection prospects for these models relative to what is shown in Fig.~\ref{fig:sensitivity}. }
\es{Enrscreen}{
\Enr > \Enr^\text{screen} \approx {1  \over 2 m_N a_0^2} \,.
}
  Additionally, we require the DM to be deflected at an angle greater than $\theta_\text{min} \sim 1$, which amounts to imposing a lower bound 
  \es{Enrhard}{
  \Enr > \Enr^\text{hard} = 2 {\mu_{N \chi}^2 \over m_N} v^2 \sin^2(\theta_\text{min} / 2) 
  }
   on the recoil energy.  The hard-scattering lower energy bound is greater than the screening bound for \mbox{$m_\chi \gtrsim 1 / (2 \sin(\theta_\text{min} / 2) a_0 v)$}.

   In the large $m_A$ scenario, there is no screening bound $E_\text{nr}^\text{screen}$ because the dark force is short range.  Similarly, in that scenario the bound $E_\text{nr}^\text{hard}$ is not necessary because the differential cross section~\eqref{dsigdEr} is independent of $E_\text{nr}$.  However, in the light mediator case the differential cross section~\eqref{diff1} rises steeply for low-angle scatters.  These low-angle scattering events do not significantly modify the trajectories of the DM, and so we exclude them by imposing the bound $E_\text{nr}^\text{hard}$ (see~\cite{Agashe:2014yua} for related discussions). 

When $\Enr^\text{hard} > \Enr^\text{screen}$, we find that 
\es{}{
{\bar \sigma_e \over \sigma_{N \chi}} = \tan^2\left(\theta_\text{min} / 2 \right) {16 v^4 \mu_{N\chi}^2 \, \mu_{e\chi}^2 \over Z^2 \,(m_e \alpha)^4} \, ,
}
while when $\Enr^\text{hard} < \Enr^\text{screen}$, 
\es{}{
{\bar \sigma_e \over \sigma_{N \chi}} = {16 \, v^4 \mu_{N\chi}^2 \, \mu_{e\chi}^2 \over Z^2 \,(m_e \alpha)^4}  \left( {1 \over 4 \mu_{N \chi}^2 a_0^2 v^2 - 1} \right)\,.
}
Requiring that there be at least one hard scatter (\mbox{$\theta_\text{min} \sim \pi/4$}) while the DM traverses the full length of the Earth's diameter  then gives the lower bound (black line) shown in the right panel of Fig.~\ref{Fig: Fq}.  

The estimates in this section are meant to roughly approximate the DM-electron scattering cross sections for which nuclear scattering in the Earth's interior becomes relevant.  As seen in Fig.~\ref{Fig: Fq}, these effects may be important over a wide range of the allowed parameter space.  The results presented here, however, require that the DM have tree-level couplings with the quarks; if this interaction is loop-suppressed, then the prospect of observing Earth effects is less optimistic.  Additionally, the current XENON10 exclusion limits are contained within the region where these Earth effects are relevant; this may mean that it is necessary to reinterpret these experimental constraints in light of the modified phase-space distribution for these particular sub-GeV DM models.  Along those lines, we have not attempted to quantify the observable consequences of the DM interactions inside the Earth.  We suspect, however, that the phenomenology could be manifested by a daily modulation of the rate.  We leave such investigations to future work.

\section{Conclusion} 

For sub-GeV DM, the main avenue for discovery in a direct-detection experiment is the detection of DM scattering off target electrons.  The properties of such a signal are affected by the inelastic nature of the electron excitation, as well as the electron ionization form factor.  We studied signals in both atomic (\emph{e.g.}, Xenon) and semiconductor (\emph{e.g.}, Germanium) targets.  In particular, we presented a new semi-analytic approach to calculate the scattering rates for semiconductor targets; this approach makes it tractable to estimate an experiment's sensitivity without relying on numerical packages to obtain detailed modeling of the semiconductor's properties.  We argue that most of the detailed band-structure physics should have little effect on the DM-electron scattering predictions.  

The annual-modulation fraction for DM-electron scattering is found to be ${\sim\mathcal{O}(10\%)}$ over a large range of candidate masses, which is somewhat higher than that expected from nuclear recoils. Observation of higher harmonic modes ($n \gtrsim 2$) is challenging, typically requiring $\sim10^{2}$--$10^{4}$ times more exposure to observe after the annual modulation. The phase of annual modulation is shown to be affected by gravitational focusing due to the Sun, similar to the case of nuclear scattering.  

We showed that DM-nuclear interactions inside the Earth can cause sub-GeV DM candidates to scatter before reaching the detector, possibly leading to a daily modulation and directional dependence of the count rate over a significant portion of the motivated parameter space. Further study is required to accurately quantify this effect, which we leave to future work.\\

\vspace{-0.35in}
\section*{Note Added}
\vspace{-0.1in}
Reference \cite{Essig:2015cda}, which appeared soon after our paper, presents a detailed numerical calculation of the DM-electron scattering rates in semiconductor targets.  The results we obtain using our semi-analytic method are in general agreement with theirs.  There are two important differences, however.  First, there is a discrepancy between the two methods in the projected limits for the case of a 1$e^-$ threshold and $F_\text{DM} \propto 1/q^2$.  We assume that the electron effective mass is equal to the free-electron mass throughout: $m_* = m_e$.  This approximation breaks down if the final-state electron is near the minimum of the conduction band, when $m_* = 0.56 \, m_e$ is more appropriate.  This correction is most relevant when low momentum recoils are enhanced and brings our results in agreement with those of~\cite{Essig:2015cda} for the case of a 1$e^-$ threshold and $F_\text{DM} \propto 1/q^2$.  Secondly, we find that the Germanium 3$d$ electrons provide a larger contribution to the scattering rate at deposited energies above $\sim$30~eV.  For this reason, our projected limits tend to be stronger than those presented in \cite{Essig:2015cda} for the 5$e^-$ and 10$e^-$ thresholds.  %Ultimately, experimental measurements will shed light on the contribution from these core electrons.

\vspace{-0.2in}
\section*{Acknowledgements}
\vspace{-0.1in}
\noindent We are especially grateful for the many enlightening discussions we have shared with Rouven Essig and Tien-Tien Yu.  We also thank Ilya Belopolski, David Huse, Christopher Malbon, Jeremy Mardon, Jesse Thaler, Tomer Volansky, and David Wallace for helpful discussions.  B.R.S was supported by a Pappalardo Fellowship in Physics at MIT.  ML is supported by the U.S. Department of Energy under grant Contract Number DE-SC0007968.

\label{sec:conc}

\newpage
\onecolumngrid
\vspace{0.3in}
\twocolumngrid
\def\bibsection{} 
\bibliographystyle{utphys}
\bibliography{DMElectronModulation}
\end{document}